\documentclass[american,aps,pra,reprint, superscriptaddress]{revtex4-1}
\usepackage{amsmath,amscd,amsthm,amssymb}
\usepackage{mathrsfs}
\usepackage{graphicx}
\usepackage{epsf,epstopdf}
\usepackage[all]{xy}
\usepackage[unicode=true,pdfusetitle, bookmarks=true,bookmarksnumbered=false,bookmarksopen=false, breaklinks=false,pdfborder={0 0 0},backref=false,colorlinks=false] {hyperref}
\hypersetup{colorlinks=true,linkcolor=myurlcolor,citecolor=myurlcolor,urlcolor=myurlcolor}
\usepackage{graphics,epstopdf,graphicx, amsthm, amsmath, amssymb, times, braket, color, bm}
\usepackage{cleveref}
\usepackage{caption}
\usepackage{subcaption}
\usepackage{makecell}

\definecolor{myurlcolor}{rgb}{0,0,0.7}

\theoremstyle{plain}

\providecommand{\theoremname}{Theorem}
\newcommand*{\myproofname}{Proof}

\makeatother

\theoremstyle{definition}

\theoremstyle{remark}

\begin{document}

\title{Matrix manipulations via unitary transformations and ancilla-state measurements}
\author{Alexander I. Zenchuk}
\email{zenchuk@itp.ac.ru}
\affiliation{Federal Research Center of Problems of Chemical Physics and Medicinal Chemistry RAS,
Chernogolovka, Moscow reg., 142432, Russia}

\author{Wentao Qi}
\email{qiwt5@mail2.sysu.edu.cn}
\affiliation{Institute of Quantum Computing and Computer Theory, School of
Computer Science and Engineering,
Sun Yat-sen University, Guangzhou 510006, China}

\author{Asutosh Kumar}
 \email{asutoshk.phys@gmail.com}
 \affiliation{Department of Physics, Gaya College, Magadh University,
Rampur, Gaya 823001, India}

\author{Junde Wu}
\email{wjd@zju.edu.cn}
\affiliation{School of Mathematical Sciences, Zhejiang University, Hangzhou 310027, China}


\begin{abstract}
We propose protocols for calculating  inner product,  matrix addition and matrix multiplication based on multiqubit Toffoli-type and the simplest one-qubit operations and employ
ancilla-measurements  to remove all garbage of calculations.
{
The depth (runtime) of addition protocol is $O(1)$ and that of
other protocols logarithmically increases with the dimensionality of the considered matrices.
}
\end{abstract}


\maketitle

{\it Introduction.--}
Quantum computers can outperform classical computers by exploiting quantum features to solve problems efficiently. Those quantum features are exploited by devising efficient quantum algorithms that take less running time (number of steps) to solve computational tasks. Quantum algorithms such as Deutsch-Jozsa, Grover’s Search, and Shor’s quantum factoring provide substantial speedup to classical algorithms.
Quantum algorithms represent a widely acknowledged area of quantum information whose intensive development is stimulated by the fast progress in constructing quantum processors based on  superconducting qubits (IBM, Google), trapped-ion technology  (ionQ), topological qubits  (Microsoft).

Finding solutions to systems of linear equations is a ubiquitous problem in science and engineering.
The Harrow-Hassidim-Loyd (HHL) algorithm \cite{HHL} is a quantum algorithm that approximates a solution to a system of linear equations with an exponential speedup over the fastest classical algorithm. Afterwards, other quantum algorithms to solve systems of linear equations were proposed \cite{CJS,BWPRWL,WZP} and some simple meaningful instances of the HHL algorithm were experimentally realised \cite{HHL4, HHL5, HHL6, HHL7}.
There is, however, a significant obstacle in realizing the control rotation of ancilla via quantum-mechanical tool in the HHL algorithm. An alternative protocol for solving systems of linear algebraic equations with a particular realization on superconducting quantum processor of IBM Quantum Experience was proposed in~\cite{SID}, which also has certain disadvantage requiring inversion of the matrix via classical algorithm.
There are many applications of the HHL-algorithm in various protocols based on matrix operations \cite{Wang,ZhaoZ, Tak}, including  solving differential equations  \cite{Berry}. {The protocols of matrix algebra proposed in \cite{ZZRF} are based on Trotterization method and Baker-Champbell-Housdorff formula for exponentiating matrices.  We underline the  relevance of quantum Fourier transform \cite{QFT1, QFT2, QFT3} and phase estimation \cite{CEMM,LP} in most of the above protocols.}
The inner product of arbitrary vectors as a matrix operation is calculated in~\cite{ZhaoZ} using an ancilla and Hadamard operator. The result is obtained via probabilistic method by performing measurements on ancilla.
There is an alternative  ``Sender-Receiver"  scheme for the inner product via a two-terminal quantum transmission line \cite{SZ_2019}.
The given vectors are encoded as the pure states of two separated senders and the result appears
in a certain element of the two-qubit receiver's {density matrix} after evolution and applying the proper unitary transformation.
This model can be modified where time-evolution is not required and matrix operations are realized using the special unitary transformations only \cite{QZKW_arxiv}.

In this paper we develop further the idea of using the unitary transformations of special type for realization of protocols of linear algebra.
We concentrate on another aspect of a matrix and  consider that its elements {are encoded into the pure  state of a quantum system.
Matrix operations (scalar product, sum and product of two matrices) are realized via unitary operations over  states of the composite quantum system supplemented with multiqubit ancilla $A$.
Then we operate a number of different quantum operations $W^{(k)}$ on the resulting states of the whole system,
and discard the garbage to obtain the required result.
First,  result $|res\rangle$ appears in  a superposition state
$|\chi\rangle = a |res\rangle+ |garb\rangle$,
$\langle \chi|\chi\rangle=1$, $\langle res|garb\rangle=0$. Stored in this way,  $|res\rangle$ can be used as an input for another protocol after discarding garbage $|garb\rangle$. Garbage can be removed by involving a one-qubit ancilla $B$ supplemented with the proper controlled projection and successive measurement on $B$ to obtain  the output $|1\rangle$ with the probability $c={|a|}\sqrt{|\langle res|res\rangle|}$,  thus mapping  $|\chi\rangle$ to  { $\frac{|res\rangle}{\sqrt{|\langle res|res\rangle|}}$}.
Throughout the paper we assume that the initial state of a quantum system is prepared in advance, although this is a problem of its own \cite{HK}.
}


{\it Inner Product.--}
We consider  two $n$-qubit subsystems  $S_1$ and $S_2$ (we set $N=2^n$). The pure states
\begin{eqnarray}
|\Psi_i\rangle =\sum_{k=0}^{N-1} a^{(i)}_{k} |k\rangle_{S_i} ,\;\;i=1,2,\;\; {\sum_{k}|a^{(i)}_{k}|^2=1},
\end{eqnarray}
encode the elements of two vectors (complex in general)
{$
a^{(i)}=(a^{(i)}_{0}\;\dots\;a^{(i)}_{N-1})^T,\;\; i=1,2,
$
}
where $|k\rangle $ is the binary representation of $k$.
Thus, each subsystem  $S_i$  is encoded into $n$ qubits and its dimensionality logarithmically increases with vector dimensionality $N$. { The initial state of the whole system is $|\Phi_0\rangle = |\Psi_1\rangle \otimes |\Psi_2\rangle$.}
We also consider an $n$-qubit ancilla $A$ in the state $|0\rangle_A$.
Now we introduce the control operators
\begin{eqnarray}
W^{(m)}_j&=&P^{(m)}_j\otimes \sigma^{(x)}_j + (I_j-P^{(m)}_j)\otimes I_{A,j} ,
\end{eqnarray}
where $P^{(m)}_j=|m_{j}\rangle_{S_1}|m_{j}\rangle_{S_2}  \;{ _{S_1}}\langle m_{j}|  {_{S_2}}\langle m_{j}| \;\; (m=1,0)$ is the projector acting on the pair of $j$th qubits of the subsystems $S_1$ and $S_2$,
$\sigma^{(x)}_j$ is the Pauli matrix, $I_{A,j}$ is the identity operator applied to the $j$th qubit of the  ancilla $A$,
$I_j$ is the 2-qubit  identity operator acting on the $j$th spins of the subsystems $S_1$ and $S_2$.   { Hereafter, in general, $I_X$ is the identity operator acting
on the system $X$}. Note that {all $W^{(m)}_j$, $m=0,1$, $j=1,\dots,n$, commute by construction}.  We apply the operator
$  W^{(1)}_{S_1S_2A} ={ \prod_{j=1}^n W^{(0)}_j W^{(1)}_j}$ on $|\Phi_0\rangle |0\rangle_A$, and obtain
{\begin{eqnarray}\label{1Phi1}
	&&|\Phi_1\rangle= W^{(1)}_{S_1S_2A}  |\Phi_0\rangle|0\rangle_A=\\\nonumber
	&&\left(\sum_{k=0}^{N-1} a^{(1)}_{k} a^{(2)}_{k} |k\rangle_{S_1}  |k\rangle_{S_2}\right) |N-1\rangle_A +
	|g_1\rangle_{S_1S_2A}.
\end{eqnarray}
}

Notice that all information needed to perform the inner product is collected in the first term of the state {$|\Phi_1 \rangle$} (\ref{1Phi1}).
The second term  $|g_1\rangle_{{S_1S_2A}}$  is the garbage which is to be eventually removed. Since all $W^{(m)}_j$ {with different  $j$} are applied to different triples of qubits, they can be applied  simultaneously.
{Now we label the result and garbage in the state $|\Phi_1\rangle$ to prevent them from mixing in the following calculations.  For this goal we introduce the
projector
$
P_A=|N-1\rangle_A\; {_A}\langle N-1|,
$
1-qubit ancilla $B_1$ in the initial state $|0\rangle_{B_1}$ and apply the control operator
$
W^{(2)}_{AB_1}=P_A\otimes \sigma^{(x)}_{B_1} +(I_{A}- P_A)\otimes  I_{B_1}
$
to the ancillae $A$ and $B_1$, respectively.
Thus we obtain
\begin{eqnarray}\nonumber
&&|\Phi_2\rangle= W^{(2)}_{AB_1}|\Phi_1\rangle |0\rangle_{B_1}=
\left(\sum_{k=0}^{N-1} a^{(1)}_{k} a^{(2)}_{k} |k\rangle_{S_1}  |k\rangle_{S_2}\right) \times \\\nonumber
&&
|N-1\rangle_A\otimes  |1\rangle_{B_1} +
|g_1\rangle_{S_1S_2A}\otimes  |0\rangle_{B_1} .
\end{eqnarray}
The control operator $W^{(2)}_{AB_1}$ with the $n$-qubit control register can be
represented in terms of $O(n)$ Toffoli operators \cite{KShV}. Therefore the depth of the circuit calculating $|\Phi_2\rangle$ is $O(n)=O(\log\,N)$.

Now we apply the Hadamard transformations $W^{(3)}_{S_1S_2A}=H^{\otimes 3n}$ to all the qubits of $|\Phi_2\rangle $ simultaneously except the ancilla $B_1$,
\begin{eqnarray}\nonumber
&&|\Phi_3\rangle =W^{(3)}_{S_1S_2A}|\Phi_2\rangle =
\frac{\langle \Psi_2^*|\Psi_1\rangle }{2^{3 n/2}}
|0\rangle_{S_1} |0\rangle_{S_2} |0\rangle_{A}|1\rangle_{B_1} +\\\nonumber
&&
 |g_3\rangle_{S_1S_2AB_1},\;\; \langle \Psi_2^*|\Psi_1\rangle=   \sum_{k=0}^{N-1} a^{(1)}_{k} a^{(2)}_{k}.
\end{eqnarray}
To label the  new garbage, we
introduce the projector
$
P_{S_1S_2AB_1}= |0\rangle_{S_1}|0\rangle_{S_2}|0\rangle_{A}|1\rangle_{B_1}\;
{_{S_1}}\langle 0| {_{S_2}}\langle 0| _{A}\langle 0|  _{B_1}\langle 1|,
$
prepare another ancilla $B_2$ in the ground state $|0\rangle_{B_2}$ and apply the control operator
$
W^{(4)}_{S_1S_2AB_1B_2}=P_{S_1S_2AB_1} \otimes \sigma^{(x)}_{B_2}+
(I_{S_1S_2AB_1} -P_{S_1S_2AB_1} )\otimes I_{B_2}\,
$
  to $|\Phi_3\rangle \otimes |0\rangle_{B_2}$:
\begin{eqnarray}\nonumber
	&&|\Phi_4\rangle =W^{(4)}_{S_1S_2AB_1B_2}|\Phi_3\rangle |0\rangle_{B_2} =\frac{\langle \Psi_2^*|\Psi_1\rangle}{2^{3 n/2}} \times
\\\nonumber
&&   |0\rangle_{S_1} |0\rangle_{S_2} |0\rangle_{A} |1\rangle_{B_1} |1\rangle_{B_2} +
	|g_2\rangle_{S_1S_2AB_1} |0\rangle_{B_2}.
\end{eqnarray}
The control operator
$W^{(4)}_{S_1S_2AB_1B_2}$ with $3n+1$ control qubits can be represented in terms of $O( 3n) = O(n)$ Toffoli gates \cite{KShV}.}

{ The inner product of two vectors is stored in a probability  amplitude. Measuring the ancilla $B_2$ with the output $|1\rangle_{B_2}$ we remove the garbage and stay with the single term in the quantum state
\begin{eqnarray}
|\Phi_5\rangle &=&\frac{\langle \Psi_2^*| \Psi_1\rangle}{|\langle \Psi_2^*| \Psi_1\rangle|  }  |0\rangle_{S_1} |0\rangle_{S_2} |0\rangle_{A}|1\rangle_{B_1},
\end{eqnarray}
which stores the phase of the inner product. The  absolute value  of the inner product is known from the probability of the above measurement which is $|\langle \Psi_2^*| \Psi_1\rangle|^2 /2^{3n}$.

The whole depth of the protocol is defined by the operators $W^{(2)}_{AB_1}$ and $W^{(4)}_{S_1S_2AB_1B_2}$, in both cases it is $O(n)=O(\log\, N)$.
The circuit is given in Fig. 1(d).
}

\begin{figure*}[!]
\centering
    \includegraphics[scale=0.6]{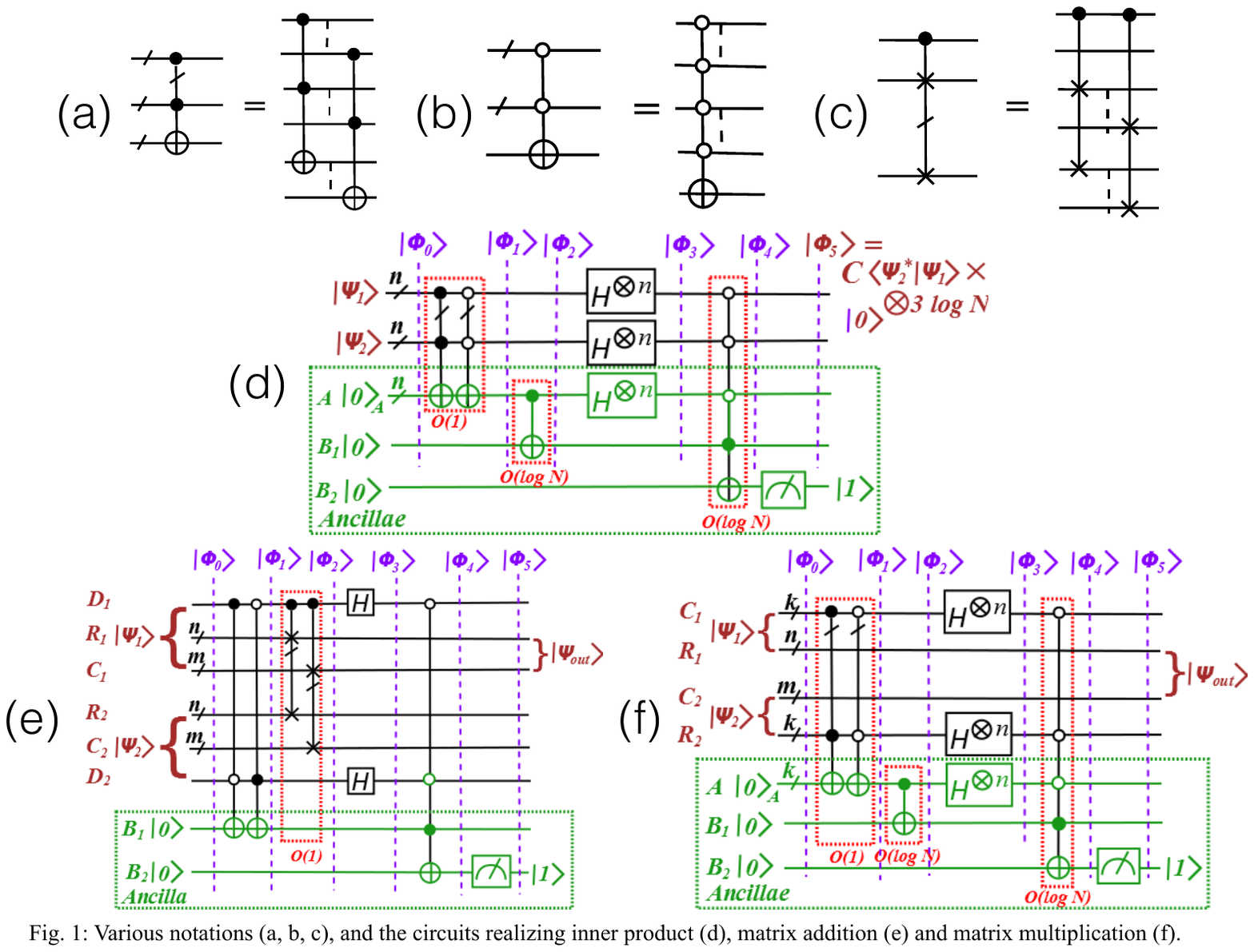}
\label{Fig:AlgOP}
\end{figure*}

{\it Matrix Addition.--}
For adding two $N\times M$ matrices  $A^{(i)}$, $i=1,2$, with the elements $\{a^{(i)}_{jk}\}$  ($N=2^n$, $M=2^m$), we first introduce two registers $R_1$ and $R_2$ of $n$ qubits and two registers $C_1$  and $C_2$ of $m$ qubits which enumerate rows and columns of both matrices, {and two additional qubits $D_1$ and $D_2$    associated with the matrices $A^{(1)}$ and $A^{(2)}$ respectively.}
The pure states encoding the elements of matrices are
\begin{eqnarray}
&&|\Psi_i\rangle = \sum_{j=0}^{N-1}  \sum_{l=0}^{M-1}   a^{(i)}_{jl} |j\rangle_{R_i} |l\rangle_{C_i} |0\rangle_{D_i} +\\\nonumber
&& s |0\rangle_{R_i} |0\rangle_{C_i} |1\rangle_{D_i} ,\;\;\sum_{jl}|a^{(i)}_{jl}|^2+|s|^2=1,\;\;i=1,2,
\end{eqnarray}
where $s$ is a parameter.
The initial state of the whole system reads
\begin{eqnarray}
&&
|\Phi_0\rangle =  |\Psi_2\rangle \otimes |\Psi_1\rangle=\\\nonumber
&&
s\sum_{j=0}^{N-1} \sum_{l=0}^{M-1}\Big( a^{(1)}_{jl}  |j\rangle_{R_1} |l\rangle_{C_1} |0\rangle_{D_1}
 |0\rangle_{R_2} |0\rangle_{C_2} |1\rangle_{D_2} +\\\nonumber
&&
a^{(2)}_{jl}  |0\rangle_{R_1} |0\rangle_{C_1} |1\rangle_{D_1}
 |j\rangle_{R_2} |l\rangle_{C_2} |0\rangle_{D_2} \Big) +
|g_1\rangle_{R_1C_1R_2C_2{ D_1D_2}} .
\end{eqnarray}
{Our aim is to organize the sum $a^{(1)}_{jl} +a^{(2)}_{jl}$ and label  the garbage.
To this end we introduce the 1-qubit  ancilla $B_1$  in the ground states $|0\rangle_{B_1}$, and define the operator
\begin{eqnarray}
W^{(m)} &=& P^{(m)}_{D_1D_2} \otimes \sigma^{(x)}_{B_1} + (I_{D_1D_2}-P^{(m)}_{D_1D_2}) \otimes I_{B_1},
\end{eqnarray}
where $\sigma^{(x)}_{B_1}$ is the Pauli matrix, and $P^{(m)}_{D_1D_2} \,\, (m=1,2)$ are the projectors
\begin{eqnarray}
P^{(1)}_{D_1D_2}&=&|1\rangle_{D_1} |0\rangle_{D_2}\; {_{D_1}}\langle 1| {_{D_2}}\langle 0|,\\\nonumber
P^{(2)}_{D_1D_2}&=&|0\rangle_{D_1} |1\rangle_{D_2}\; {_{D_1}}\langle 0| {_{D_2}}\langle 1|.
\end{eqnarray}
{Obviously, $[W^{(1)},W^{(2)}]=0$.}
Applying the  operator $W^{(1)}_{D_1D_2B_1}=W^{(1)}W^{(2)}$ to $|\Phi_0\rangle |0\rangle_{B_1}$ we obtain:
\begin{eqnarray}
&&
|\Phi_1\rangle =W^{(1)}_{D_1D_2B_1}    |\Phi_0\rangle \otimes
|0\rangle_{B}=\\\nonumber
&&
s \sum_{j=0}^{N-1}  \sum_{l=0}^{M-1} \Big( a^{(1)}_{jl}  |j\rangle_{R_1} |l\rangle_{C_1} |0\rangle_{D_1}
 |0\rangle_{R_2} |0\rangle_{C_2} |1\rangle_{D_2}+\\\nonumber
&&
 a^{(2)}_{jl}  |0\rangle_{R_1} |0\rangle_{C_1} |1\rangle_{D_1}
 |j\rangle_{R_2} |l\rangle_{C_2} |0\rangle_{D_2}\Big) |1\rangle_{B_1}+ \\\nonumber
 &&
 |g_1\rangle_{R_1C_1R_2C_2D_1D_2} |0\rangle_{B_1}.
\end{eqnarray}
Now we construct {the control operator}
\begin{eqnarray}\nonumber
&&
W^{(2)}_{D_1R_1C_1R_2C_2}=|1\rangle_{D_1} \;{_{D_1}}\langle 1| \otimes SWAP_{R_1,R_2}SWAP_{C_1,C_2} + \\\nonumber
&&
|0\rangle_{D_1} \;{_{D_1}}\langle 0| \otimes
I_{R_1C_1R_2C_2}
\end{eqnarray}
that acts on $|\Phi_1\rangle$ and swaps the states of $R_1$ with $R_2$ and states of $C_1$with $C_2$ to yield
\begin{eqnarray}
&&
|\Phi_2\rangle =W^{(2)}_{D_1R_1C_1R_2C_2}|\Phi_1\rangle =\\\nonumber
&&
s \sum_{j=1}^N  \sum_{l=1}^M \Big( a^{(1)}_{jl}  |j\rangle_{R_1} |l\rangle_{C_1} |0\rangle_{D_1}
 |0\rangle_{R_2} |0\rangle_{C_2} |1\rangle_{D_2}+\\\nonumber
&&
 a^{(2)}_{jl}  |j\rangle_{R_1} |l\rangle_{C_1} |1\rangle_{D_1}
 |0\rangle_{R_2} |0\rangle_{C_2} |0\rangle_{D_2}\Big)|1\rangle_{B_1} +\\\nonumber
 &&
  |g_2\rangle_{R_1C_1R_2C_2D_1D_2} |0\rangle_{B_1}.
\end{eqnarray}
{We notice that the SWAPs in the control operator $W^{(2)}_{D_1R_1C_1R_2C_2}$ have common single control and are related  to different pairs of qubits; therefore they can be applied simultaneously.  Consequently, the depth of this operator    is $O(1)$.}
Next, we apply the Hadamard operators $W^{(3)}_{D_1D_2}=H_{D_1}H_{D_2}$ to $D_1$ and $D_2$:
\begin{eqnarray}
&&|\Phi_3\rangle =W^{(3)}_{D_1D_2} |\Phi_2\rangle =
\frac{s}{2}\sum_{j=0}^{N-1}  \sum_{l=0}^{M-1}  (a^{(1)}_{jl}+  \\\nonumber
&&
a^{(2)}_{jl})  |j\rangle_{R_1} |l\rangle_{C_1} |0\rangle_{D_1}
 |0\rangle_{R_2} |0\rangle_{C_2} |0\rangle_{D_2}|1\rangle_{B_1} +
\\\nonumber
	&&
  |g_3\rangle_{R_1C_1D_1R_2C_2D_2B_1}.
\end{eqnarray}
Thus, the sum of two matrices is stored in the first term of $|\Phi_3\rangle$.
To label the  garbage,  we prepare the 1-qubit ancilla $B_2$ in the state $|0\rangle_{B_2}$,
introduce the projector
$
P_{D_1,D_2,B_1}=  |0\rangle_{D_1} |0\rangle_{D_2}|1\rangle_{B_1} \; {_{D_1}}\langle 0| {_{D_2}} \langle 0| _{B_1}\langle 1|
$ and apply the control operator
$
W^{(4)}_{D_1D_2B_1B_2} =P_{D_1D_2B_1} \otimes \sigma^x_{B_2} + (I_{D_1D_2B_1}-P_{D_1D_2B_1})\otimes I_{B_2}\,
$
 to $|\Phi_3\rangle |0\rangle_{B_2}$:
{\begin{eqnarray}\nonumber
	&&|\Phi_4\rangle= W^{(4)}_{D_1D_2B_1B_2}|\Phi_3\rangle |0\rangle_{B_2}=
	\frac{s}{2}\sum_{j=0}^{N-1}  \sum_{l=0}^{M-1} \Big( (a^{(1)}_{jl} + \\\nonumber
	&&
	a^{(2)}_{jl})  {|j\rangle_{R_1} |l\rangle_{C_1} |0\rangle_{D_1}
		|0\rangle_{R_2} |0\rangle_{C_2} |0\rangle_{D_2}}\Big) |1\rangle_{B_1}  |1\rangle_{B_2}+ \\\nonumber
	&&
	|g_2\rangle_{R_1C_1D_1R_2C_2D_2B_1} |0\rangle_{B_2}.
\end{eqnarray}
}
Finally, on measuring the ancilla $B_2$ with the output  $|1\rangle_{B_2}$ we remove the garbage and obtain
\begin{eqnarray}\nonumber
&&
|\Phi_5\rangle=
|\Psi_{out}\rangle |0\rangle_{D_1}
 |0\rangle_{R_2} |0\rangle_{C_2} |0\rangle_{D_2}|1\rangle_{B_1},\\\nonumber
 &&
|\Psi_{out}\rangle =G^{{ -1}} \sum_{j=0}^{N-1}  \sum_{l=0}^{M-1}   (a^{(1)}_{jl}+a^{(2)}_{jl})   |j\rangle_{R_1} |l\rangle_{C_1},
\end{eqnarray}
where the normalization $G{=(\sum_{jl} |a^{(1)}_{jl}+a^{(2)}_{jl}|^2)^{1/2}}$ is known from the probability of the above measurement
which is {$s^2G^2/4$}.  { It follows from the  above consideration   that the depth of this  protocol is $O(1)$.}
The circuit  is given in Fig. 1(e).
}

{\it Matrix Multiplication.--}
We present a protocol for multiplying  $N\times K$ matrix $A^{(1)}$ by $K\times M$  matrix $A^{(2)}$,  with the elements $A^{(i)}=\{a^{(i)}_{jk}\}$, $i=1,2$,
{ assuming $N = 2^n$, $K = 2^k$, $M = 2^m$ with positive
integers $n$, $k$, $m$.}

We first introduce one  register of {$n$} qubits, two registers of   {$k$} qubits and one register of {$m$} qubits which enumerate rows and columns of both matrices.
The pure states encoding the elements of matrices are
\begin{eqnarray}\label{inst}
|\Psi_i\rangle &=& \sum_{jl}  a^{(i)}_{jl} |j\rangle_{R_i} |l\rangle_{C_i},\\\label{norm}
&&\sum_{jl} |a^{(i)}_{jl}|^2=1.
\end{eqnarray}
The initial state of the whole system reads
\begin{eqnarray}
&&|\Phi_0\rangle =  |\Psi_2\rangle \otimes |\Psi_1\rangle=\\\nonumber
&&\sum_{j_1=0}^{N-1}\sum_{l_1,j_2=0}^{K-1}\sum_{l_2=0}^{M-1}
a^{(1)}_{j_1l_1}a^{(2)}_{j_2l_2}  |j_1\rangle_{R_1} |l_1\rangle_{C_1} |j_2\rangle_{R_2} |l_2\rangle_{C_2}.
\end{eqnarray}
We also consider the $k$-qubit  ancilla $A$  in the ground state $|0\rangle_{A}$.
Now we define the operators $W^{(m)}_{j} \;\; (m=0,1)$
{\begin{eqnarray}
W^{(m)}_{j} &=& P^{(m)}_j \otimes \sigma^{(x)}_{A,j} + (I_{j}-P^{(m)}_{j}) \otimes I_{A,j},
\end{eqnarray}
}
where $I_{A,j}$  is the identity operators acting on the $j$th qubit of the ancilla $A$,
$I_j$ is the identity operator acting on the 2-qubit subsystem including the $j$th qubits of $C_1$ and $R_2$, and $P^{(m)}_j = |m_j\rangle_{C_1}|m_j\rangle_{R_2}\; {_{C_1}}\langle m_j| {_{R_2}}\langle m_j|$ are the projectors acting on the $j$th cubits of $C_1$ and $R_2$.
All operators $W^{(m)}_j$, $m_j=0,1$, $j=1,\dots, K$ commute with each other.
Applying the operator $W^{(1)}_{C_1R_2A}={\prod_{j=1}^k W^{(1)}_j W^{(0)}_{j}}$   to $|\Phi_0\rangle |0\rangle_A$ we obtain:
\begin{eqnarray}
&&
|\Phi_1\rangle =W^{(1)}_{C_1R_2A} |\Phi_0\rangle
|0\rangle_{A}=\\\nonumber
&&\left({\sum_{j_1=0}^{N-1}}\sum_{j=0}^{K-1}\sum_{l_1=0}^{M-1}
a^{(1)}_{j_1 j}a^{(2)}_{j l_1}  |j_1\rangle_{R_1} |j\rangle_{C_1} |j\rangle_{R_2} |l_1\rangle_{C_2}\right) |{K-1}\rangle_A +\\\nonumber
&&|g_1\rangle_{R_1C_1R_2C_2A}  .
\end{eqnarray}
Since the operators $W^{(1)}_j$ and $W^{(0)}_{j}$ with different $j$ are applied to different triples of qubits, they can be performed in parallel.
{To label the garbage, we introduce the projector
$
P_{A}=|K-1\rangle_A \; {_A}\langle K-1|
$
together with the 1-qubit ancilla $B_1$ in the ground state $|0\rangle_{B_1}$.
Then we construct the control  operator
$
W^{(2)}_{AB_1}=P_A\otimes \sigma^{(x)}_{B_1} + (I_A-P_A)\otimes I_{B_1},\,
$
and apply it to $|\Phi_1\rangle |0\rangle_{B_1}$:
\begin{eqnarray}
&&
|\Phi_2\rangle= W^{(2)}_{AB_1} |\Phi_1\rangle |0\rangle_{B_1} = \\\nonumber
&&
\left({\sum_{j_1=0}^{N-1}\sum_{j=0}^{K-1}\sum_{l_1=0}^{M-1}}
a^{(1)}_{j_1 j}a^{(2)}_{j l_1}  |j_1\rangle_{R_1} |j\rangle_{C_1} |j\rangle_{R_2} |l_1\rangle_{C_2}\right) \times \\\nonumber
&&
|K-1\rangle_A |1\rangle_{B_1} +
|g_1\rangle_{R_1C_1R_2C_2A} |0\rangle_{B_1}.
\end{eqnarray}
This {$k$}-qubit control  operator has depth $O(k)$.
Now we apply the Hadamard transformations $W^{(3)}_{C_1R_2A}=H^{\otimes 3 k}$  to $C_1$, $R_2$ and $A$:
\begin{eqnarray}\nonumber
&&
|\Phi_3\rangle =W^{(3)}_{C_1R_2A}  |\Phi_2\rangle
 =\\\nonumber
&&
\frac{1}{2^{3 k/2}}\left({\sum_{j_1=0}^{N-1}}\sum_{j=0}^{K-1}\sum_{l_1=0}^{M-1}
a^{(1)}_{j_1 j}a^{(2)}_{j l_1}  |j_1\rangle_{R_1} |0\rangle_{C_1} |0\rangle_{R_2} |l_1\rangle_{C_2}\right) \times \\\nonumber
&& |0\rangle_A |1\rangle_{B_1}+
|g_2\rangle_{R_1C_1R_2C_2AB_1}.
\end{eqnarray}
Here the first term contains the desired matrix product.
Next, to label the new garbage, we prepare another one-qubit ancilla $B_2$ in the ground state $|0\rangle_{B_2}$, introduce the
projector
$
P_{C_1R_2{A}B_1} = |0\rangle_{C_1}|0\rangle_{R_2} {|0\rangle_A}|1\rangle_{B_1}
 \;{_{C_1}}\langle 0| {_{R_2}}\langle 0| _A\langle 0|_{B_1}\langle 1|
$
and the  control operator
$
W^{(4)}_{C_1 R_2 {A} B_1B_2} = P_{C_1R_2 {A}B_1} \otimes \sigma^{(x)}_{B_2} + (I_{C_1R_2 {A}B_1}- P_{C_1R_2 {A}B_1})\otimes I_{B_2}\,
$ of the depth $O(k)$ with $({3}k+1)$-qubit control register.
Applying this operator  to $|\Psi_3\rangle |0\rangle_{B_2}$ we obtain
\begin{eqnarray}
&&
|\Phi_4\rangle = W^{(4)}_{C_1 R_2 {A} B_1B_2}|\Phi_3\rangle |1\rangle_{B_2} =\\\nonumber
&&
{\frac{1}{2^{3 k/2}}}\left({\sum_{j_1=0}^{N-1}}\sum_{j=0}^{K-1}\sum_{l_1=0}^{M-1}
a^{(1)}_{j_1 j}a^{(2)}_{j l_1}  |j_1\rangle_{R_1} |0\rangle_{C_1} |0\rangle_{R_2} |l_1\rangle_{C_2}\right)\times\\\nonumber
&&  |0\rangle_A|1\rangle_{B_1} |1\rangle_{B_2}+ |g_3\rangle_{R_1C_1R_2C_2AB_1} |0\rangle_{B_2}.
\end{eqnarray}
Performing measurement over $B_2$ with the output $|1\rangle_{B_2}$ we remove garbage and obtain
\begin{eqnarray}\nonumber
&&
|\Phi_5\rangle =
|\Psi_{out}\rangle\, |0\rangle_{C_1} |0\rangle_{R_2}  |0\rangle_A|1\rangle_{B_1} ,
\\\nonumber
&&
|\Psi_{out}\rangle =G^{{-1}} {\sum_{j_1=0}^{N-1}}\sum_{j=0}^{K-1}\sum_{l_1=0}^{M-1}
a^{(1)}_{j_1 j}a^{(2)}_{j l_1}  |j_1\rangle_{R_1} |l_1\rangle_{C_2},
\end{eqnarray}
where the normalization $G{=(\sum_{j_1,l_2} |\sum_j a^{(1)}_{j_1 j}a^{(2)}_{j l_1}|^2 )^{1/2}}$ is known from the probability of the above measurement {which} equals $G^2/2^{3k}$.
The result of multiplication is stored in the registers  $R_1$ and $C_2$.
From the above analysis we conclude that the depth  of the  whole  protocol is defined by the operators $W^{(2)}_{AB_1}$  and $W^{(4)}_{C_1R_2{A}B_1B_2}$ and equals  $O(k)=O(\log(K))$.
The circuit  is given in Fig. 1(f).}

{We  emphasize that inner vector  product
 and matrix addition  can be recast as matrix multiplication.} The inner product of two $N$-element vectors  is the product of $1\times N$ and $N\times 1$ matrices $A^{(1)}$ and $A^{(2)}$, while the sum of $N\times M$ matrices $A^{(1)}$ and $A^{(2)}$ can be found in the  result of the product of the following $2N \times 2 M$ matrices
\begin{eqnarray}\nonumber
&&
\tilde A^{(1)} =\left(
\begin{array}{cc}
A^{(1)} & I_{NM}\cr
0_{NM}& 0_{NM}
\end{array}
\right),\;\;
\tilde A^{(2)} =\left(
\begin{array}{cc}
I_{NM}& 0_{NM}\cr
A^{(2)}& 0_{NM}
\end{array}
\right)\;\; \Rightarrow \\\nonumber
&&\tilde A^{(1)} \tilde A^{(2)} = \left(
\begin{array}{cc}
A^{(1)}+A^{(2)}& 0_{NM}\cr
0_{NM}& 0_{NM}
\end{array}
\right),
\end{eqnarray}
where $I_{NM}$ and $0_{NM}$ are, respectively, the $N\times M$ identity and zero matrices.

{\it Remark on probability amplification.}
In the algorithms of calculating the inner product and matrix multiplication,
the probability of obtaining the needed ancilla state $|1\rangle$ in result of  measurement is not large, it is $\sim 1/N^3\le 1/2$. Partially, the problem of small probability can be solved performing the set of $L$ experiments on different processors, although this method is not very effective in our case because the probability of needed result in single measurement doesn't exceed  $1/2$. For instance,
we assume that the state   $|0\rangle$ of the ancilla appears with probability $1-1/N^3$ in result of the measurement. Then, performing $L=N^3$ experiments we obtain that the probability of getting  $|0\rangle$ in all experiments is $(1-1/N^3)^{N^3}$ tends to $ e^{-1}$ as $N\to \infty$. Then the probability of measuring $|1\rangle$ is $(1-e^{-1})\to  0.632$.
This is rather large value, but the price is the increase in  the required  space  $N^3$ times.

{\it Example of matrix multiplication.}
As an example, we multiply two $2\times 2$ matrices
\begin{eqnarray}
A_1=\left(\begin{array}{cc}
0.4&0.4\cr
0.2&0.8
\end{array}\right),\;\;A_2=\left(\begin{array}{cc}
0.4&0.2\cr
0.4&0.8
\end{array}\right).
\end{eqnarray}
 Thus, $N=M=K=2$, $n=m=k=1$. Normalizations (13) hold for  these matrices.
 Each subsystem $R_i$, $C_i$, $i=1,2$, includes only one qubit and
 \begin{eqnarray}
 |\Psi_1\rangle &=& 0.4|0\rangle_{R_1}|0\rangle_{C_1} + 0.4|0\rangle_{R_1}|1\rangle_{C_1} +\\\nonumber
 &&0.2|1\rangle_{R_1}|0\rangle_{C_1} + 0.8|1\rangle_{R_1}|1\rangle_{C_1},\\\nonumber
 |\Psi_2\rangle &=& 0.4|0\rangle_{R_2}|0\rangle_{C_2} + 0.2|0\rangle_{R_2}|1\rangle_{C_2} +\\\nonumber
 &&0.4|1\rangle_{R_2}|0\rangle_{C_2} + 0.8|1\rangle_{R_2}|1\rangle_{C_2}.
 \end{eqnarray}
 The ancilla $A$ includes one qubit,
operators $W^{m)}_1$ are given by the expression
\begin{eqnarray}
W^{(m)}_1= P^{(m)}_1 \otimes \sigma^{(x)}_{A} + (I_1-P_1^{(m)})\otimes I_{A},\;\; m=0,1,
\end{eqnarray}
where projectors read
\begin{eqnarray}
&&
P^{(0)}_1= |0\rangle_{C_1}|0\rangle_{R_2} \; _{C1}\langle 0| _{R_2}\langle 0|, \\\nonumber
&&P^{(1)}_1= |1\rangle_{C_1}|1\rangle_{R_2} \; _{C1}\langle 1| _{R_2}\langle 1|.
\end{eqnarray}
We also have
\begin{eqnarray}
W^{(2)}_{AB_1}=P_A\otimes \sigma^{(x)}_{B_1} + (I_A-P_A)\otimes I_{B_1}
\end{eqnarray}
with $P_A=|1\rangle_A \;_A\langle 1|$. The operator
$W^{(3)}_{C_1R_2A}=H^{\otimes 3}$, the projector
$P_{C_1R_2AB_1}$ remains the same as well as the operator $W^{(4)}_{C_1R_2AB_1B_2}$.
Finally, we obtain after  measurement of the ancilla $B_2$ resulting in   $|1\rangle_{B_2}$  with the probability  $G^2/2^3=0.1106$, $G=\sqrt{0.8848}$:
\begin{eqnarray}
|\Psi_{out}\rangle &=& G^{-1} \Big( 0.32    | 0\rangle_{R_1} | 0 \rangle_{C_2}  +  0.4 | 0\rangle_{R_1} | 1 \rangle_{C_2}  + \\\nonumber
&& 0.4 | 1\rangle_{R_1} | 0 \rangle_{C_2}  + 0.68  | 1\rangle_{R_1} | 1 \rangle_{C_2} \Big).
\end{eqnarray}

{\it Conclusion.--}
We proposed protocols for inner product of two vectors, matrix addition and matrix multiplication.
The protocols employ tensor product of quantum states to get product of matrix elements, the Hadamard transformations convert those products into sums, and ancilla measurements remove the garbage that appear along with the useful result.
In all three protocols the result is conserved in the probability amplitudes of certain quantum states, so that the matrices  obtained as result of multiplication or addition can be used in further calculations. {It is remarkable that the depth of the protocols  for inner product  and matrix multiplication increases logarithmically with the dimension of the considered matrices, while that of addition  protocol is $O(1)$ and doesn't depend on matrix dimensionality. }
{
The depth (runtime) of addition protocol is $O(1)$ and that of
other protocols logarithmically increases with the dimensionality of the considered matrices.
}

{\it Acknowledgments.--}
The authors thank referee for  important suggestions and remarks.
This project is supported by the National Natural Science Foundation of
China (Grants No. 12031004, No. 12271474 and No. 61877054), the
Fundamental Research Foundation for the Central Universities (Project No.
K20210337),
and Zhejiang University Global Partnership Fund 188170+194452119/003.
The work was partially funded by a state task of Russian Fundamental
Investigations (State Registration No. 124013000760-0).

\end{document}